\documentclass{emulateapj}
\usepackage{ulem}
\usepackage{longtable}
\usepackage{graphicx}
\usepackage{txfonts}
\usepackage{colortbl}
\usepackage{multirow}

\def\ergs{\ensuremath{\rm{\, erg \,s^{-1} }}}

\begin{document}
\title{Relationship Between the Kinetic Power and Bolometric Luminosity of Jets: limitation from black hole X-ray binaries, active galactic nuclei, and gamma-ray bursts}
\shorttitle{Correlation between Jet Kinetic Power and Intrinsic Jet Luminosity}
\shortauthors{Ma R., Xie F. G., \& Hou S.}

\author{Renyi Ma$^1$, Fu-Guo Xie$^2$, Shujin Hou$^{1,3}$}
\shortauthors{Ma R., Xie F. G., \& Hou S.}
\altaffiltext{1}{Department of Astronomy and Institute of Theoretical Physics and Astrophysics, Xiamen University, Xiamen 361005, Fujian Province, China; ryma@xmu.edu.cn}
\altaffiltext{2}{Key Laboratory for Research in Galaxies and Cosmology, Shanghai Astronomical Observatory, Chinese Academy of Sciences, 80 Nandan Road, Shanghai 200030, China; fgxie@shao.ac.cn}
\altaffiltext{3}{Purple Mountain Observatory, Chinese Academy of Sciences, Nanjing 210008, China; houshujin@stu.xmu.edu.cn}

\begin{abstract}
The correlation between the kinetic power $P_{\rm jet}$ and intrinsic bolometric luminosity $L_{\rm jet}$  of jets may reveal the underlying jet physics in various black hole systems.
Based on the recent work by \citet{Nemmen12}, we re-investigate this correlation with additional sources of black hole X-ray binaries (BXBs) in hard/quiescent states and low-luminosity active galactic nuclei (LLAGNs).
The new sample includes 29 sets of data from 7 BXBs and 20 LLAGNs, with $P_{\rm jet}$ and $L_{\rm jet}$ being derived from spectral modeling of the quasi-simultaneous multi-band spectra under the accretion-jet scenario.
Compared to previous works, the range of luminosity is now enlarged to more than $20$ decades, i.e. from $\sim 10^{31}\ergs$ to $\sim 10^{52}\ergs$, which allows for better constraining of the correlation.
One notable result is that the jets in BXBs and LLAGNs almost follow the same $P_{\rm jet} - L_{\rm jet}$ correlation that was obtained from blazars and gamma-ray bursts (GRBs).
The slope indices we derived are $1.03\pm0.01$ for the whole sample, $0.85\pm0.06$ for the BXB subsample, $0.71\pm0.11$ for the LLAGN subsample, and $1.01\pm0.05$ for the LLAGN-blazar subsample, respectively.
The correlation index around unit implies the independence of jet efficiency on the luminosity or kinetic power.
Our results may further support the hypothesis that similar physical processes exist in the jets of various black hole systems.
\end{abstract}


\section{Introduction}
Powerful, highly collimated jets are common in various astrophysical black hole systems, e.g. BXBs, GRBs and active galactic nuclei (AGNs). These jets originate from the vicinity of black holes--presumably from the innermost accretion disks or the central black holes, and propagate a long distance before interacting with the circumambient medium.
The jets are still an enigma and play important roles in astrophysics, for example, in AGN feedback (\citealt{Fabian12}).

Although the jets in AGNs and GRBs are diverse in their bulk velocities, mass loss rates $\dot{M}_{\rm jet}$, and spectra etc., they are believed to have the same nature \citep[e.g.][]{Mirabel10}.
A recent work by \citet{Wang11} showed that GRB afterglows and blazars have a similar relation between the radio luminosity and spectral slope in radio to optical bands.
\citet{Wu11} pointed out that the luminosity and Doppler factor are correlated in a unified form for GRBs and blazars.
Very recently, \citet[][hereafter N12]{Nemmen12} studied a sample of 234 blazars and 54 GRBs with relatively good constraints on both jet kinetic power and intrinsic jet bolometric luminosity \footnote{For blazars and GRBs, this is equivalent to the $\gamma$-ray luminosity because the radiation is dominated by this band and the contribution from accretion disk is negligible due to the beaming effects of the jet.}, and discovered a tight correlation, $P_{\rm jet}\approx 4.6\times10^{47} (L_{\rm jet}/10^{47}\ergs)^{0.98}\ergs$.
\citet{Zhang13} found a tighter relation between the jet power and intrinsic synchrotron peak luminosity by systematically modeling the spectral energy distributions for blazars and ¦Ã-ray narrow-line Seyfert 1s, but the slope is changed to $0.79\pm0.01$.

An interesting question is whether the jets in BXBs and LLAGNs also follow the correlation of N12, or in other words, are jets in BXBs and LLAGNs similar to those in blazars and GRBs?
However, this is not an easy question to answer because of the great difficulties in measuring the kinetic power and intrinsic luminosity of jets in these sources.

For AGNs, different methods have been postulated to estimate the jet power,
such as model-independent measurements of the X-ray cavity or radio bubbles (e.g., \citealt{Godfrey13} and references therein) and the empirical relationships between the jet power and quantities such as specific/bolometric radio luminosity, strength of line emission, and Bondi rate, etc. \citep{Falle91,Kaiser97,Willott99,Allen06,Gu09,Wu11a}.
But for BXBs, only very few sources have been resolved with cavities \citep{Gallo05, Hao09}, and the jet power is poorly constrained.

Moreover, the intrinsic luminosity of jets in BXBs and LLAGNs are also hard to constrain.
Unlike GRBs and blazars, in which the radiation is dominated by jets due to beaming effects, BXBs and LLAGNs suffer contamination from the emission of accretion disks.

In this Letter, we try to estimate the jet power and intrinsic luminosity by modeling multi-band spectra.
Theoretically, a detailed spectrum can well constrain the physical conditions of the jet, and consequently, the jet power and luminosity.
Although this method is model-dependent, the result would be reliable if the model is well-proved.
Here we use the coupled accretion-jet model \citep{Yuan05a} to fit the quasi-simultaneous multi-band spectra of LLAGNs and BXBs in hard/quiescent states, and then calculate the jet power and luminosity.
Our method is the same to \citet{Zhang13} except the different model.

This Letter is organized as follows. In Section 2 the coupled accretion jet model is briefly described, focusing on the jet model we adopted. The sample is introduced in Section 3, and the main results are given in Section 4. A short discussion and summary are given in Sections 5 and 6, respectively.

\section{Accretion-Jet Model for BXBs and LLAGNs}
Only a brief description of the coupled accretion-jet model is given here, and \citet[][hereafter Y05]{Yuan05a} is referred for details. This model includes three components, i.e., a standard thin disk which is truncated at a certain radius, a hot accretion flow within such a radius, and a relativistic jet.
For the hot accretion flow, we take into account the advances in our understanding over the past decade, namely, the existence of outflow and the viscous heating to electrons (see \citealt{Xie12} and references therein). Roughly speaking, the thin disk, the hot accretion flow, and the jet mainly contribute to the radiation in the infrared-ultraviolet, X-ray, and radio bands, respectively. The jet may dominate the X-ray emission when the source is extremely faint \citep{Yuan05b}.

This model provides a comprehensive explanation for the broadband spectra of LLAGNs and BXBs in hard/quiescent states (see \citealt{Yuan07b} for a review), and the timing features observed in the BXB system of XTE J1118+480 (Y05).
In addition to the standard correlation between radio and X-ray luminosity, the model predicts that the radio/X-ray correlation steepens when the X-ray luminosity is lower than a critical value \citep{Yuan05b},
which has been confirmed by observations of LLAGNs \citep{Pellegrini07,Wrobel08,deGasperin11,Yuan09}.
These successes provide confidence on the jet power and intrinsic luminosity we derive.

\subsection{Jet Model}
Since the power and luminosity of jets are the focus of this work, we provide a little more description of the jet model as follows.
Our jet model is phenomenological without considering the driving process.
The jet is composed of normal plasma, i.e., electrons and protons.
The plasma in the jet moves outward relativistically with the bulk Lorentz factor $\Gamma$ and half-opening angle $\theta_j=0.1$.
Within the jet itself, different shells of the moving plasma are assumed to have different random velocities.
When the faster but later shells catch up with the slower and earlier ones, internal shocks occur.
From diffusive shock acceleration theory, a fraction (typically $1\%$) of the thermal electrons will be accelerated into a power law energy distribution.
The power law index $p_{\rm e}$ is typically constrained to be $2<p_{\rm e}<2.4$.
Because of the efficient radiative cooling of the most energetic electrons, the steady-state energy distribution of these non-thermal electrons is a broken power law.
The index changes to $p_{\rm e}+1$ above a certain energy, which can be determined self-consistently.
Following the widely adopted approach in the study of GRBs, the energy density of accelerated electrons and amplified magnetic fields are determined by two parameters, $\epsilon_e$ and $\epsilon_B$, which describe the fractions of the shock energy that go into power law electrons and magnetic fields, respectively.

Because of the existence of magnetic fields, the power law electrons will emit synchrotron photons, which dominate in the radio band and possibly in the X-ray band.
The jet is assume to be compact and extend to about $10^8 GM/c^2$, which is in the kpc scale for AGNs and the AU scale for BXBs.

\begin{table*}[!th]
\centering
\vspace{-0.3cm}
\caption{\large Data of Black-hole X-ray Binaries}
\begin{tabular}{lclcccccccl}
\hline
\hline
  \colhead{Source}
& \colhead{Distance}
& \colhead{$M_{\rm BH}$}
& \colhead{$\dot M_{\rm jet}$}
& \colhead{$\theta$}
& \colhead{$\epsilon_e$}
& \colhead{$\epsilon_B$}
& \colhead{$p_{\rm e}$}
& \colhead{Log $L_{\rm jet}$}
& \colhead{Log $P_{\rm jet}$}
& \colhead{Reference}\\
& \colhead{(kpc)}
& \colhead{($M_\odot$)}
& \colhead{($\dot M_{\rm Edd}$)}
& \colhead{(deg.)}
& & & & \colhead{($\ergs$)}
& \colhead{($\ergs$)}
& \\
\hline
XTE J1118+480       & 1.8 & 8.0  & $3.0 \times 10^{-4}$ & 70 & 0.06 & 0.02 & 2.24 & 35.67 & 35.78 & 1 \\  
                    &     & 7.2  & $7.2 \times 10^{-8}$ & 70 & 0.06 & 0.02 & 2.24 & 31.29 & 32.11 & 2 \\
XTE J1550-564       & 5.3 & 10.5 & $4.0 \times 10^{-3}$ & 70 & 0.06 & 0.02 & 2.23 & 37.10 & 37.02 & 3 \\  
                    & 5.3 & 9.6  & $2.9 \times 10^{-6}$ & 73 & 0.06 & 0.02 & 2.20  & 33.40 & 33.84 & 4 \\  
V404 Cyg            & 3.5 & 11.7 & $3.4 \times 10^{-6}$ & 56 & 0.06 & 0.02 & 2.24 & 33.75 & 33.99 & 4 \\  
SWIFT J1753.5-0127  & 6.0 & 9.0  & $9.6 \times 10^{-5}$ & 63 & 0.04 & 0.02 & 2.10 & 34.86 & 35.33 & 5 \\  
GRO J1655-40        & 3.2 & 6.3  & $2.8 \times 10^{-6}$ & 85 & 0.06 & 0.02 & 2.24 & 33.21 & 33.65 & 4 \\  
XTE J1720-318       & 8.0 & 5.0  & $7.2 \times 10^{-5}$ & 60 & 0.06 & 0.08 & 2.10 & 34.79 & 34.95 & 5 \\  
IGR J17177-3658     & $\sim$ 25 & 10.0&$9.6\times 10^{-4}$& 70 & 0.06 & 0.02 & 2.25 &36.42 & 36.38 & 6 \\   
\hline
\end{tabular}

\tablerefs{
(1) Y05;
(2) \citet{Yuan05b};
(3) \citet{Yuan07a};
(4) \citet{Pszota08};
(5) \citet{Zhang10};
(6) \citet{Ma12}.
}
\vspace{-0.5cm}
\end{table*}

\begin{table*}[!th]
\centering
\caption{\large Data of Low-luminosity AGNs}
\begin{tabular}{llclcccllccccl}
\hline
\hline
  \colhead{Source}
& \colhead{$z$ or}
& \colhead{$M_{\rm BH}$}
& \colhead{$F_{\rm 1.4GHz} ^a$}
& \colhead{$\dot M_{\rm jet}$}
& \colhead{$\theta$}
& \colhead{$\Gamma$}
& \colhead{$\epsilon_e$}
& \colhead{$\epsilon_B$}
& \colhead{$p_{\rm e}$}
& \colhead{Log $L_{\rm jet}$}
& \colhead{Log $P_{\rm jet}$}
& \colhead{Log $P'_{\rm jet}$ $^b$}
& \colhead{Reference}\\
& \colhead{{$D_L$ (Mpc)$^c$}}
& \colhead{($M_\odot$)}
& \colhead{(Jy)}
& \colhead{($\dot M_{\rm Edd}$)}
& \colhead{(deg.)}
& & & & & \colhead{($\ergs$)}
& \colhead{($\ergs$)}
& \colhead{($\ergs$)}
& \\
\hline
3C 66B   & 0.021 & $6.9 \times 10^8$ & 10.1  & $1.2 \times 10^{-4}$ & 45 & 2.3 & 0.05 & 0.01 & 2.4  & 44.71 & 44.11 & 44.77 &  1,6 \\
3C 317   & 0.035 & $6.3 \times 10^8$ & 5.46   & $3.9 \times 10^{-5}$ & 50 & 2.3 & 0.2  & 0.15 & 2.25 & 44.00 & 43.60 & 44.90 &  2,6 \\
3C 346   & 0.162 & $7.7 \times 10^8$ & 3.7  & $1.6 \times 10^{-4}$ & 25 & 2.3 & 0.05 & 0.01 & 2.4  & 45.22 & 44.31 & 45.80 &  1,6 \\
3C 449   & 0.017 & $6.9 \times 10^8$ & 3.69  & $4.6 \times 10^{-5}$ & 70 & 2.3 & 0.1 & 0.005 & 2.4  & 44.41 & 43.71 & 44.30 &  2,6 \\
NGC 266  & 0.016 & $4.0 \times 10^7$ & 0.0106  & $1.2 \times 10^{-5}$ & 30 & 2.3 & 0.1  & 0.01 & 2.5  & 42.81 & 41.87 & 42.35 &  3,8 \\
NGC 383  & 0.017 & $8.5 \times 10^7$ & 4.8  & $6.9 \times 10^{-5}$ & 45 & 2.3 & 0.2  & 0.02 & 2.4  & 43.97 & 42.98 & 44.39 &  4,6 \\
NGC 2484 & 0.043 & $8.5 \times 10^8$ & 2.22  & $4.1 \times 10^{-5}$ & 34 & 2.3 & 0.12 & 0.01 & 2.23  & 44.63 & 43.75 & 44.74 &  2,6 \\
\hline
NGC 1052 & 17.8  & $1.3 \times 10^8$ & 0.9132  & $8.0 \times 10^{-4}$ & 60 & 10  & 0.2  & 0.02 & 2.3  & 46.11 & 45.07 & 42.92 &  3,7 \\
NGC 3031 & 1.4   & $6.3 \times 10^7$ & 0.55  & $1.4 \times 10^{-6}$ & 50 & 2.3 & 0.1  & 0.01 & 2.2  & 41.60 & 41.15 & 41.10 &  5,8 \\
NGC 3169 & 19.7  & $6.3 \times 10^7$ & 0.0892    & $5.8\times 10^{-6}$  & 30 & 2.3 & 0.01 & 0.01 & 2.2  & 41.49 & 41.77 & 42.23 & 3,8 \\
NGC 3226 & 23.4  & $1.3 \times 10^8$ &  0.0032    & $1.4\times 10^{-6}$  & 30 & 2.3 & 0.1  & 0.01 & 2.3  & 42.10 & 41.47 & 41.26 & 4,8 \\
NGC 3368 & 8.1   & $2.5 \times 10^7$ &  0.0316   & $8.0\times 10^{-5}$  & 60 & 10  & 0.1  & 0.01 & 2.3  & 43.80 & 43.35 & 41.32 & 3,7 \\
NGC 3998 & 21.6  & $7.9 \times 10^8$ & 0.1014  & $7.4 \times 10^{-6}$ & 30 & 2.3 & 0.01 & 0.001& 2.2  & 42.38 & 42.98 & 42.33 &  3,8 \\
NGC 4203 & 9.7   & $1.0 \times 10^7$ &  0.0061    & $1.5\times 10^{-5}$  & 25 & 10  & 0.1  & 0.02 & 2.2  & 42.69 & 42.23 & 40.90 & 4,7 \\
NGC 4261 & 35.1  & $5.0 \times 10^8$ &  18.6  & $1.6 \times 10^{-5}$ & 63 & 2.3 & 0.1  & 0.01 & 2.2  & 43.65 & 43.12 & 44.35 &  1,8 \\
NGC 4374 & 16.8  & $7.9 \times 10^8$ &  6.1  & $9.2 \times 10^{-7}$ & 30 & 2.3 & 0.01 & 0.1  & 2.4  & 41.70 & 42.08 & 43.51 &  1,8 \\
NGC 4486 & 16.8  & $6.3 \times 10^9$ & 210.0  & $1.8 \times 10^{-7}$ & 10 & 6.0 & 0.001 & 0.008 & 2.3& 41.60 & 42.85 & 44.66 &  4,8 \\
NGC 4552 & 16.8  & $1.6 \times 10^8$ & 0.100  & $4.6 \times 10^{-6}$ & 30 & 2.3  & 0.1 & 0.1 & 2.6   & 42.48 & 42.08 & 42.17 &  4,8 \\
NGC 4579 & 16.8  & $6.3 \times 10^7$ & 0.0982  & $3.2 \times 10^{-5}$ & 45 & 2.3 & 0.011 & 0.008& 2.2  & 41.81 & 42.52 & 42.16 &  3,8 \\
NGC 4594 & 20.0  & $3.2 \times 10^8$ &  0.094   & $1.8\times 10^{-5}$  & 30 & 2.3 & 0.005& 0.003& 2.3  & 42.15 & 42.98 & 42.26 & 3,8 \\
\hline\\
\end{tabular}
\tablecomments{
$^a$ the radio flux at 1.4~GHz are taken from NED: \url{http://ned.ipac.caltech.edu}. The observations of medium magnitude with uncertainty given are selected, and the references are listed in the last column;
$^b$ jet kinetic power derived from radio luminosity at 1.4 GHz;
$^c$ The redshifts are taken from simbad: \url{http://simbad.u-strasbg.fr/simbad/sim-fid}, and the luminosity distance of the sources are taken from \citet{Ho09}.}

\tablerefs{
(1)\citet{s1};
(2)\citet{s9};
(3)\citet{s3};
(4)\citet{s4};
(5)\citet{s8};
(6)\citet{Wu07};
(7)\citet{Yu11};
(8)\citet{Nemmen11}.
}
\vspace{0.5cm}
\end{table*}

\subsection{Bolometric Luminosity and Bulk Kinetic Power of Jet}
Multi-band (from radio up to X-ray) spectra are used to constrain the modeling parameters.
The contribution of the jet, the hot accretion flow, and the thin disk can be separated by spectral modeling, as well as the parameters, such as the bulk Lorentz factor $\Gamma$, the viewing angle of the jet $\theta$, and the mass loss rate $\dot{M}_{\rm jet}$.
The jet power can be obtained as
\begin{equation}
P_{\rm jet}=(\Gamma-1)\dot{M}_{\rm jet}c^2.
\end{equation}
It should be noted that $\dot{M}_{\rm jet}$ is defined as $\Gamma$ times the value in the cited paper.
Despite the differences in the definition of $\dot{M}_{\rm jet}$, the final jet power is the same  \citep[e.g.][]{Pszota08}.
As for the intrinsic luminosity, $L_{\rm jet}$,which indicates the total luminosity of the jet,
it is calculated by integrating the observed intensity over different inclination angles.
For GRBs and blazars, the Lorentz factors are large, $\Gamma \ga 10$, the beaming effects are significant,
and almost all of the jet emission is concentrated in a small cone with a half-opening angle of $\theta_j$.
In this case, $L_{\rm jet}$ can be simplified as the product of isotropic luminosity and beaming factor $1-\cos\theta_j$, as done in N12.
However, for BXBs and LLAGNs, the beaming effects are not significant, and the observed intensity should be integrated over different inclination angles.

It is difficult to measure $\Gamma$ from observations, and we simply take the typical value.
For jets of BXBs in their hard or quiescent states, previous analyses indicate that the Lorentz factors should be relatively small, i.e. $\Gamma\lesssim1.67$ \citep{Gallo03}, and we adopt $\Gamma=1.2$. For jets in AGNs, the Lorentz factors have larger scatters, and we adopt the value from the cited paper (see Table 2 for details).

\section{Sample}
By searching the literature, sources that have been well-modeled with the coupled accretion-jet model are collected according to the following criteria.
First, the modeling parameter should be reasonable.
For example, due to energy conservation, the value of $\epsilon_e$ and $\epsilon_B$ cannot be too large.
They are also unlikely to be very small, otherwise the radiative efficiency would be too low to be observed.
Because $\epsilon_e \sim \sqrt{\epsilon_B}$ \citep{Medvedev06} and $\epsilon_e < 1$, $\epsilon_B$ should be less than $\epsilon_e$.
Considering the uncertainty in diffusive shock acceleration theory,
$10^{-4}<\epsilon_e,\epsilon_B <0.5$, and $\epsilon_B < 10 \epsilon_e$ are taken.
Second, because the radio emission is dominated by jet,
the source should have at least two radio data points.
This criterion is relaxed in the case when the X-ray spectra can only be explained by a jet model.
Third, only the modeling with the most typical parameters is selected if the source has been modeled more than once for a given observation.
Finally we collect 7 BXBs and 20 LLAGNs, in total 29 sets of observations and modeling, as shown in Tables 1 and 2.

In order to reduce the systematic uncertainties of the model, only observations that have been investigated with the same accretion-jet model are included.
The jet kinetic power and luminosity are by products of spectral modeling, without considering the $P_{\rm jet} - L_{\rm jet}$ correlation reported by N12.

The distances are taken from the \citet{Ho09} or calculate from the redshift $z$ with cosmological constants of
$H_0 = 70\ {\rm km ~s^{-1} ~Mpc^{-1}}, \Omega_M = 0.27, \Omega_\Lambda = 0.73$.
Considering that the distances used in previous modeling are a little different, we re-calculate the luminosity and re-model the spectra by slightly adjusting $\dot{M}_{\rm jet}$.
The 1 $\sigma$ uncertainties of the derived $P_{\rm jet}$ and $L_{\rm jet}$ are taken to be $0.2$ dex for BXBs and $0.5$ dex for LLAGNs.

\section{Results}

\begin{figure*}
\centering
\includegraphics[width=12cm]{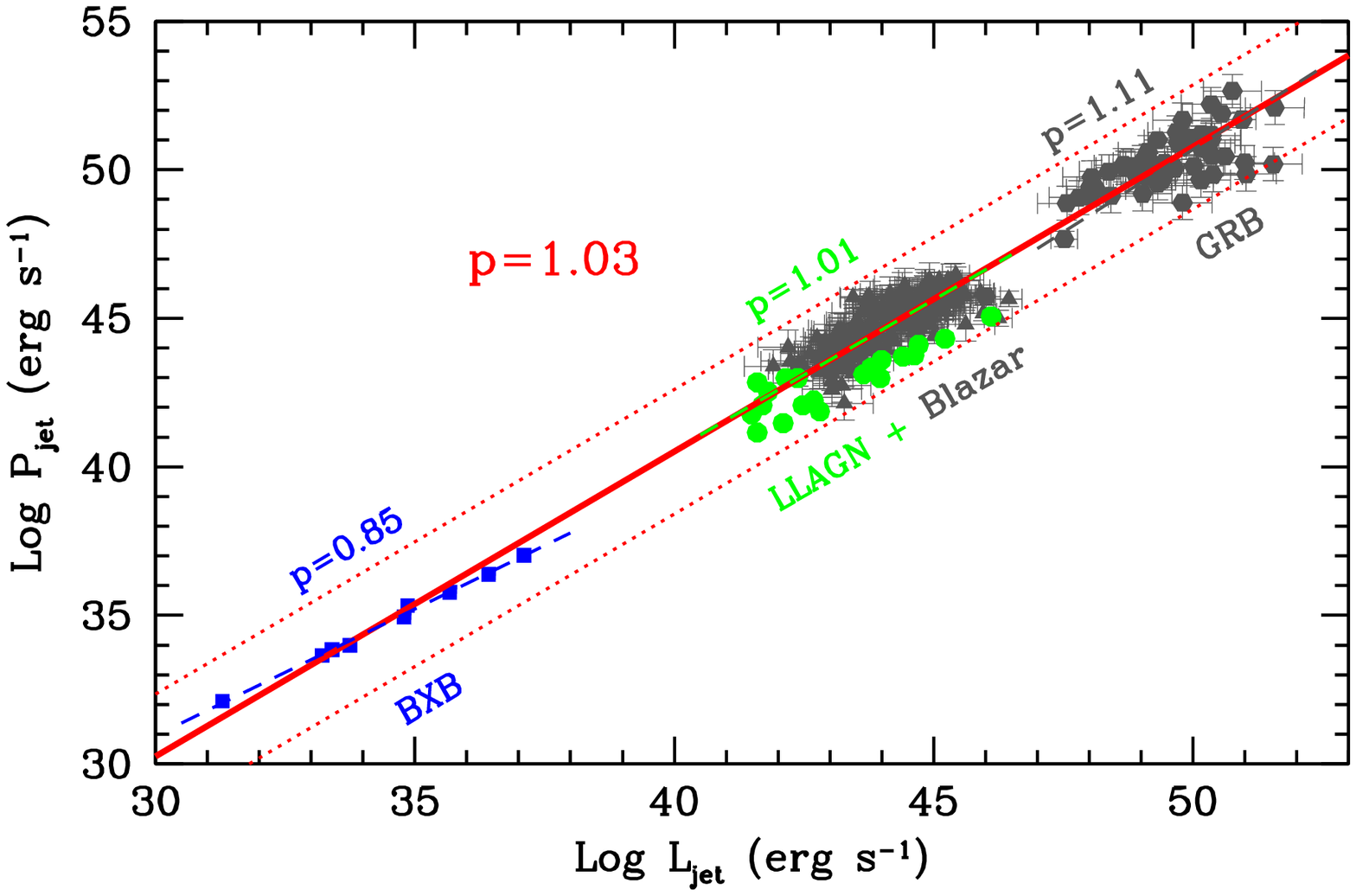}
\caption{Relationship between the kinetic power $P_{\rm jet}$ and the intrinsic jet luminosity $L_{\rm jet}$. The green circles and blue squares are the data for LLAGNs and BXBs, respectively. The gray hexagons and triangles represent data for GRBs and blazars, taken from N12. The thick red solid line is the fitting result of the whole sample (see Equation\ \ref{eq:newfit}), while the two red dotted lines indicate the 3-$\sigma$ deviation. The dashed curves (marked with slope indices) with different colors show the fitting results of each subsample.}
\end{figure*}

Figure 1  shows the correlation for the whole sample including our new data and those from N12.
It can be seen that the jet power and the intrinsic jet luminosity are correlated and can be well-fitted with the form $P_{\rm jet}\propto L_{\rm jet}^p$.
For BXBs and LLAGNs, the fitting formula are
\begin{eqnarray}
P_{\rm jet,BXB} &=& 10^{35.08\pm3.89}\left(L_{\rm jet,BXB}/10^{35} \ergs \right)^{0.85\pm0.06}\ergs,\\
P_{\rm jet,LLAGN} &=& 10^{42.85\pm9.33}\left(L_{\rm jet,LLAGN}/10^{43} \ergs \right)^{0.71\pm0.11}\ergs.
\end{eqnarray}
The errorbars here and below indicate the 1-$\sigma$ uncertainties of the fits  . Since both LLAGNs and blazars belong to the AGN category and they overlap in the intrinsic luminosity (as shown in Fig. 1), we fit the LLAGN and blazar subsamples jointly. The fitting result of this AGN sample is,
\begin{equation}
P_{\rm jet,AGN} = 10^{43.60\pm 4.26}\left(L_{\rm jet,AGN}/10^{43} \ergs \right)^{1.01\pm0.05}\ergs.
\end{equation}
We also investigate the correlation for subsamples of blazars and GRBs and the slope indices are found to be $p=0.89\pm0.05$ and $1.11\pm0.08$, respectively.
Combining the above results, the slope indices of different subsamples are all $\sim 0.9$, which are coherent within error ranges.
This supports the hypothesis that the jets in each subsample may share similar physical processes.
The whole sample is therefore fitted as
\begin{equation}
P_{\rm jet} = 10^{48.91\pm0.85}\left(L_{\rm jet}/10^{48} \ergs \right)^{1.03\pm0.01}\ergs,
\label{eq:newfit}
\end{equation}
with the Pearson correlation coefficient of $0.95$. We note that this correlation is consistent with N12 and agrees with \citet{Zhang13} within uncertainty.

Considering the consistent $P_{\rm jet}$-$L_{\rm jet}$ correlation, the larger luminosity range is very helpful in accurately determining the correlation index.
With the data of the BXBs, the range of luminosity increases from $10$ orders of magnitude in N12 to more than $20$ orders of magnitude, i.e. from $\sim10^{31}\ergs$ to $\sim10^{52}\ergs$.
Obviously, further investigation with better constraints on $P_{\rm jet}$ and $L_{\rm jet}$ is needed to better understand the correlation.

\section{Discussion}

Although there is a lack of direct evidence, we speculate that this correlation is real and intrinsic. First, the slope of each subsample is consistent with that of the whole sample. Second,  differences in distance can not be the cause, since BXBs are all within our Galaxy and follow the same tight correlation. Third, differences in black hole mass can not be responsible for this correlation, since both BXBs and GRBs are stellar systems and follow the same correlation. Fourth, the mass-loss rate of jet $\dot{M}_{\rm jet}$ is still not the cause, since $\dot{M}_{\rm jet}$ (in unit $\dot{M}_{\rm Edd}$) overlap for our BXB+LLAGN subsample.

This correlation may be partly explained by the jet model \citep{Heinz03}.
For compact jet with flat radio spectra in BXBs and LLAGNs, the luminosity is dominated by radio-infrared emission, and the correlation index is about 0.7 \citep{Coriat11}, which is roughly consistent with our results.

\subsection{Calculation of the Jet Power}

As mentioned in Section 1, there are many ways to estimate the jet kinetic power for AGNs, of which measurements from X-ray cavities are currently the most reliable \citep[e.g.][]{Birzan04,Cavagnolo10}.
We collect the 1.4 GHz radio data and calculate the jet power with the fitting formula given in \citet{Cavagnolo10}, as also shown in Table 2.
It can be seen that the results by spectral modeling usually agree well with those from empirical formula.
The corresponding $P_{\rm jet}-L_{\rm jet}$ relations for LLAGNs are shown in Figure 2.
The slope is $0.83\pm0.18$ for jet power calculated from radio emission, which is quite similar to our result $0.71\pm0.11$.
The similarities show the validity of the spectral modeling method in estimation of jet kinetic power.

Moreover, three of our LLAGNs (NGC 4374, NGC 4486, and NGC 4552) show well-resolved X-ray cavities by {\it Chandra}, and their jet power has been derived to be about $5 \times10^{42}\ergs$, $1\times10^{43}\ergs$, $5\times10^{41}\ergs$, respectively \citep{Allen06,Cavagnolo10}.
Our jet powers for these three sources agree well with these results.

\begin{figure}
\begin{center}
\includegraphics[width=0.8\columnwidth]{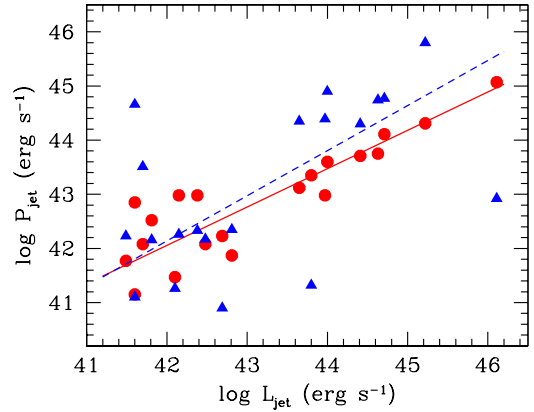}
\caption{Correlation between the jet luminosity and kinetic power. The jet luminosity is derived by modeling the multi-band spectra, while the kinetic power is derived from the spectral modeling (red solid circle) and radio luminosity at 1.4 GHz (blue triangle).
The red solid and blue dashed lines are the corresponding linear fit to the circles and triangles, respectively.}
\end{center}
\end{figure}

\subsection{Radiative Efficiency of the Jet}

As shown in Figure 3, we investigate the radiative efficiency of the jet, which is estimated as (N12; \citealt{Zhang07}) $\varepsilon_{\rm rad}=L_{\rm jet}/(L_{\rm jet}+P_{\rm jet})$. Here $L_{\rm jet}+P_{\rm jet}$ is the total power, ignoring the power accelerating non-thermal protons. Substitute Equation~\ref{eq:newfit}, the ``averaged'' radiative efficiency of jets can be written as
\begin{equation}
\varepsilon_{\rm rad} \approx 1/\left[1+8.13(L_{\rm jet}/10^{48}\ergs)^{0.03}\right].\label{eq:eff}
\end{equation}
From this equation, it can be seen that the efficiency is almost independent of jet luminosity.
Despite this general trend, the sources, even within each subsample, are diverse in their $\varepsilon_{\rm rad}$ (see also Figure 3 in N12).

\begin{figure}
\begin{center}
\includegraphics[width=0.8\columnwidth]{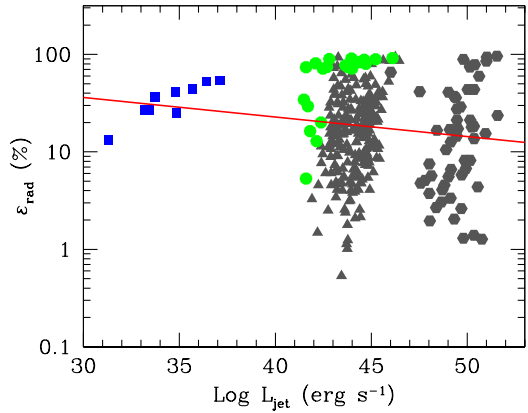}
\caption{Relationship between the radiative efficiency $\varepsilon_{\rm rad}$ and intrinsic bolometric luminosity $L_{\rm jet}$ of the jet. The symbols for the data points are the same as to those in Figure 1. The thick, red, solid curve corresponds to the efficiency of Equation\ \ref{eq:eff}.}
\end{center}
\end{figure}

Whether there is a weak dependence of jet radiative efficiency on jet luminosity or kinetic power is worth further investigation.
This is probably due to the combined effects of acceleration and emission of non-thermal particles.
For a given shock, the acceleration efficiency is affected by the shock strength, magnetic field, and the pre-existing non-thermal particles \citep[e.g.][]{Kang07}.
Although the bulk Lorentz factor may not affect the efficiency for a given shock strength \citep{Bykov11}, jets with larger Lorentz factor may produce more and/or stronger internal shocks, which make the particle acceleration more efficient.
However, the radiative efficiency of jets with larger bulk Lorentz factor may decrease to due to annihilation of high energy photons.

\section{summary}
In this Letter we extend the work of N12 by including observational data from low-luminosity systems, i.e., LLAGNs and BXBs in the hard/quiescent states. The kinetic power and luminosity of jets in these systems are determined through a spectral modeling method. With our new sample, the dynamical range in luminosity, from $\sim 10^{31}~\ergs$ to $\sim 10^{52}~\ergs$, is now significantly enlarged.
It is found that, despite the huge differences in the jets of BXBs, LLAGNs, blazars, and GRBs, they follow an universal correlation, with the slope of each subsample being $p\approx 1.03$.

In order to understand the underlying physics of jets, further investigations of this correlation are obviously needed. In contrast to the extensive simultaneous multi-band observations of BXBs, the observations with detailed spectral modeling for the accretion-jet model are still very limited.
Furthermore, some BXBs have undergone frequent outbursts covering a sufficiently large range of $L_{\rm jet}$.
As simultaneous multi-band observations keep accumulating, it may be possible for us to study this correlation for an individual BXB. With the lowest possible uncertainties (in distance, black hole mass, and black hole spin, etc.), such work should be unique.
The influence of black hole spin on efficiency is another issue worth addressing. For the current sample of BXBs, the highest spin is $0.7\pm0.1$ for GRO 1655-40, which is not high enough to illustrate the role of spin.
Moreover, it is unclear whether the jets in other systems like neutron stars, follow the same correlation.

\acknowledgments
We are grateful to Feng Yuan, Rodrigo Nemmen and the anonymous referee for constructive suggestions and comments. We thank Xue-Feng Wu, Wei-Hua Lei and Liang Chen for informative discussions.
This work is supported in part by the Natural Science Foundation of China (grants 11333004, 11203057, 11133005, 11233006, 11103059, 11121062), the CAS/SAFEA International Partnership Program for Creative Research Teams, and the Natural Science Foundation of Fujian Province (No. 2011J01023). \\

\end{document}